\documentclass[aps,twocolumn,epsfig,rotate,showpacs]{revtex4}
\usepackage{epsfig}
\usepackage{graphicx}
\usepackage{amsmath}

\begin{document} 
\title{Directed anomalous diffusion without a biased field: A ratchet
accelerator}
\author{Jiangbin Gong \cite{bylineone} and Paul Brumer}
\affiliation{Chemical Physics Theory Group, Department of Chemistry,
University of Toronto, 
Toronto, Canada M5S 3H6}
\date{\today}

\begin{abstract}
Directed classical current that increases linearly with time without using a biased external field
is obtained
in a simple model Hamiltonian system derived from a
modified
kicked rotor model, by breaking the spatial symmetry
of the transporting regular islands in the classical phase space. A parallel
study of the corresponding quantum dynamics suggests that although quantum
coherence effects suppress the directed current and can even induce a
current reversal,
directed
quantum current that increases linearly with time
can nonetheless be realized in a quantum system that is far from the classical limit.
\end{abstract}

\maketitle

\section{introduction}
Ratchet effects, i.e., directed transport without a biased field
due to a broken spatial-temporal symmetry, are of great
interest to the understanding of molecular motors in biological systems and
have been under intense investigation in a variety of simple model systems \cite{reimann02,astumian}.
Early studies focused on the role of external noise, which was then replaced
by deterministic chaotic dynamics with dissipation \cite{jung96,mateos,porto,flach}. Recently, it was shown
that even purely Hamiltonian dynamics with both regular and chaotic phase space
structures \cite{schanz00,schanz01,cheon02,denisov02} or with complete chaos \cite{flach,monteriro02,jonckheere03,jones03}
is capable of generating ratchet effects.
As such,
there has been a great interest in studying simple model Hamiltonian systems 
in an effort to observe and understand ratchet effects.

The kicked rotor (KR) model \cite{casatibook}  and its variants are ideal systems
for such studies. For example, delta-kicked systems with an alternating kicking
period  and an asymmetric potential \cite{cheon02,monteriro02,jonckheere03},
or with multiple external fields \cite{denisov02}
have been considered in connection with ratchet effects.  These studies are stimulating since
the classical dynamics of the KR (i.e., the standard map)
and the associated quantum dynamics
have long served as paradigms for classical and quantum chaos
\cite{casatibook}. Indeed, the KR has been realized in several atom optics experimental groups \cite{raizenetc}, 
and is also of
considerable interest in other fields such as condensed
matter physics \cite{fishmanprl,casatiprl}, molecular physics
\cite{fishman,averbukh,machholm,averbukh2}, and quantum information  science
\cite{facchi,georgeot}.

Although not often thought of as such, the basic idea that underlies
the above-mentioned studies of ratchet effects in the KR,
i.e., to manipulate symmetry properties to induce
a  directed current, is not new from a coherent control perspective.
Indeed, fifteen years ago
Kurizki, Shapiro, and Brumer \cite{kurizki} proposed a simple 
coherent control scenario that can generate
photocurrents in semiconductors without a biased voltage.  This proposal and its 
various extensions \cite{hache,bhat} have been
experimentally realized \cite{dupont,hache2,stevens}.

In this paper we introduce a totally different type of Hamiltonian ratchet effect 
in a simple delta-kicked model system, which we call
a ``ratchet accelerator". Without a net external force (i.e., the external force averaged over space
is zero),  the ratchet accelerator
can generate 
directed classical anomalous transport, with the net current (defined below)
that accelerates linearly with time.
This is made possible by taking advantage of the large transporting islands in the classical phase space of a modified kicked rotor system recently
proposed by Gong, W\"{o}rner, and Brumer \cite{gongpre1,gongpre2},
in conjunction with a second kicking field that breaks the spatial symmetry.  
The corresponding quantum dynamics is also studied, shedding considerable
light on quantum-classical differences and correspondence
in the presence of anomalous diffusion. For
example,  we show that due to the tunneling between a chaotic sea and
a relatively large regular region of the classical phase space,  the quantum directed current
can be in the opposite direction of the classical directed current.

This paper is organized as follows. In Sec. II we describe the
background and the motivation of our model
as a ratchet accelerator.
The classical results of directed
anomalous diffusion are
presented
in Sec. III, followed by the analogous quantum results
in Sec. IV.  Concluding remarks are given in Sec. V.

\section{The model system}
Consider the following model Hamiltonian
\begin{eqnarray}
H(p,x,t)&=&p^{2}/2m+\lambda_{1} \cos(x/x_{0})\sum_{n}f(n)\delta(t/T-n) \nonumber \\
&& - \lambda_{2} \sin(2x/x_{0})\sum_{n}\delta(t/T-n),
\label{ratchetH}
\end{eqnarray}
where
$p$ is the momentum, $m$ is the particle mass,  $x$ is
the conjugate position,  $\lambda_{1}$ and  $\lambda_{2}$
are the strength of the first and second kicking fields, $T$ is the time interval between kicks, and the function
$f(n)$ will be determined below.
The associated classical map is given by
\begin{eqnarray}
\tilde{p}_{N} &= &\tilde{p}_{N-1}+\kappa_{1}f(N)\sin(\tilde{x}_{N-1})+2\kappa_{2}\cos(2 \tilde{x}_{N-1}); \nonumber \\
\tilde{x}_{N}&=&\tilde{x}_{N-1}+ \tilde{p}_{N},
\label{clamapp}
\end{eqnarray}
where
$\tilde{x}=x/x^{0}$ is the scaled dimensionless position variable,
$\tilde{p}\equiv p T/(mx_{0})$ is the scaled dimensionless momentum variable, $\kappa_{1}=\lambda_{1}T^{2}/(mx_{0}^{2})$, $\kappa_{2}=\lambda_{2}T^{2}/(mx_{0}^{2})$,
and $(\tilde{p}_{N}, \tilde{x}_{N})$ represents
the phase space location of a classical
trajectory at $(N+1-0^{+})T$.
The corresponding quantum unitary evolution operator 
for propagating from time $(N-0^{+})T$ to time  $(N+1-0^{+})T$ is given
by
\begin{eqnarray}
\hat{F}=\exp\left[i\frac{\tau}{2}\frac{\partial^{2}}{\partial \tilde{x}^{2}}\right]
\exp[-if(N)k_{1}\cos(\tilde{x})]\exp[ik_{2}\sin(2\tilde{x})],
\label{qmapkr}
\end{eqnarray}
with dimensionless parameters
$ k_{1}=\lambda_{1} T/\hbar$, $ k_{2}=\lambda_{2} T/\hbar$, $\tau=\hbar T/m x_{0}^{2}$.
Let
$\hat{\tilde{p}}\equiv \hat{p} T/(mx_{0})$, where $\hat{p}=-i\hbar\partial/\partial x$,
then one has $[\tilde{x},\hat{\tilde{p}}]=i\tau$, and $\hat{\tilde{p}}=-i\tau \partial/\partial \tilde{x}$,
indicating that
$\tau$ plays the role of an effective Planck constant.
Note that $\kappa_{1}=k_{1}\tau$ and $\kappa_{2}=k_{2}\tau$.

Clearly, if $\lambda_{2}=0$ and $f(n)=1$ then Eq. (\ref{ratchetH})
reduces to the standard kicked rotor model \cite{casatibook} and the classical map
of Eq. (\ref{clamapp}) becomes the standard map. If $\lambda_{2}=0$, $f(n)=g(n)$, where $g(n)=1$ for $n=4j+1$, $4j+2$, and  $g(n)=-1$ for
$n=4j+3$, $4j+4$, where $j$ an integer, then the system becomes 
that in our previous work \cite{gongpre1,gongpre2}.  Our
accelerator, introduced here, assumes
$\lambda_{2}\ne 0$ and
$f(n)=g(n)$. That is, the system is kicked by two fields, 
one of which reverses the sign of its kicking potential after every two kicks.

Note first the existence of ``accelerating trajectories'' in the
classical delta-kicked dynamics.
In particular, for particular values of $\kappa_{1}$, the standard map [$\kappa_{2}=0$, $f(n)=1$]
can generate
trajectories whose momentum increases (or decreases) linearly with time (at least on the average).  These trajectories
are a class of {\it transporting}  trajectories \cite{schanz01}.  To see this
consider
the initial conditions: $
(\tilde{p}=2\pi l_{1}, \tilde{x}=\pm \pi/2)$
for $\kappa=2\pi l_{2}$, where $l_{1}$ and $l_{2}$ are integers.  Clearly, these phase space points
are shifted by a constant value ($\pm 2\pi l_{2}$) in $\tilde{p}$ after each iteration,
resulting in a
quadratic
increase of rotational energy.  These accelerating trajectories
are rather stable insofar as they may persist
for values of $\kappa$ close to $2\pi l_{2}$ (with their average momentum
shift after each iteration oscillating around the constant value $\pm 2\pi l_{2}$),
thus giving rise to
transporting regular islands \cite{schanz01}, also called the ``accelerator modes'' in the standard map case
\cite{zaslavsky02,grigolini}.
Dramatically, if $f(n)=g(n)$ and $\kappa_{2}=0$,
then there can exist much larger 
transporting islands \cite{gongpre1}. These islands are associated with the marginally stable points
$(\tilde{p}=(2l_{1}+1)\pi, \tilde{x}=\pm \pi/2)$ for  $\kappa=(2l_{2}+1)\pi$.
Trajectories on these islands
will be shifted by a constant value [$\pm (2l_{2}+1)\pi$] in $\tilde{p}$
after each kick.

If classical
trajectories are launched from the transporting regular islands, they
simply jump to other similar islands located in adjacent phase space cells.
For trajectories initially outside the transporting regular islands,
the ``stickiness'' of the boundary between
the islands and the chaotic sea induces
anomalous diffusion over the phase space. For example,
the square of the variance of momentum increases
nonlinearly, but not quadratically.  This is intrinsically different from the
case of normal chaotic diffusion in which the  square of the variance of momentum increases linearly with the number of
kicks.

As in previous studies on delta-kicked systems,  one needs to break the spatial symmetry of the kicking potential
to induce ratchet effects.  More specifically,  
our choice of the above model Hamiltonian is motivated by the desire to have large transporting islands
in the classical phase space
\cite{gongpre1} and to break the spatial symmetry of the  transporting regular islands.  
As shown below, the symmetry can be indeed broken by a second kicking field, and yet the remaining transporting islands 
are still of a significant size and have unidirectional transporting
properties.  This being the case, the mechanism of the Hamiltonian ratchet effects described below becomes transparent
without using
previous analyses of the relationship between broken spatial-temporal symmetries and ratchet effects
\cite{flach,denisov,denisov2,denisov3}.

\section{Classical results}
\begin{figure}[ht]
\begin{center}
\vspace{-5cm}
\epsfig{file=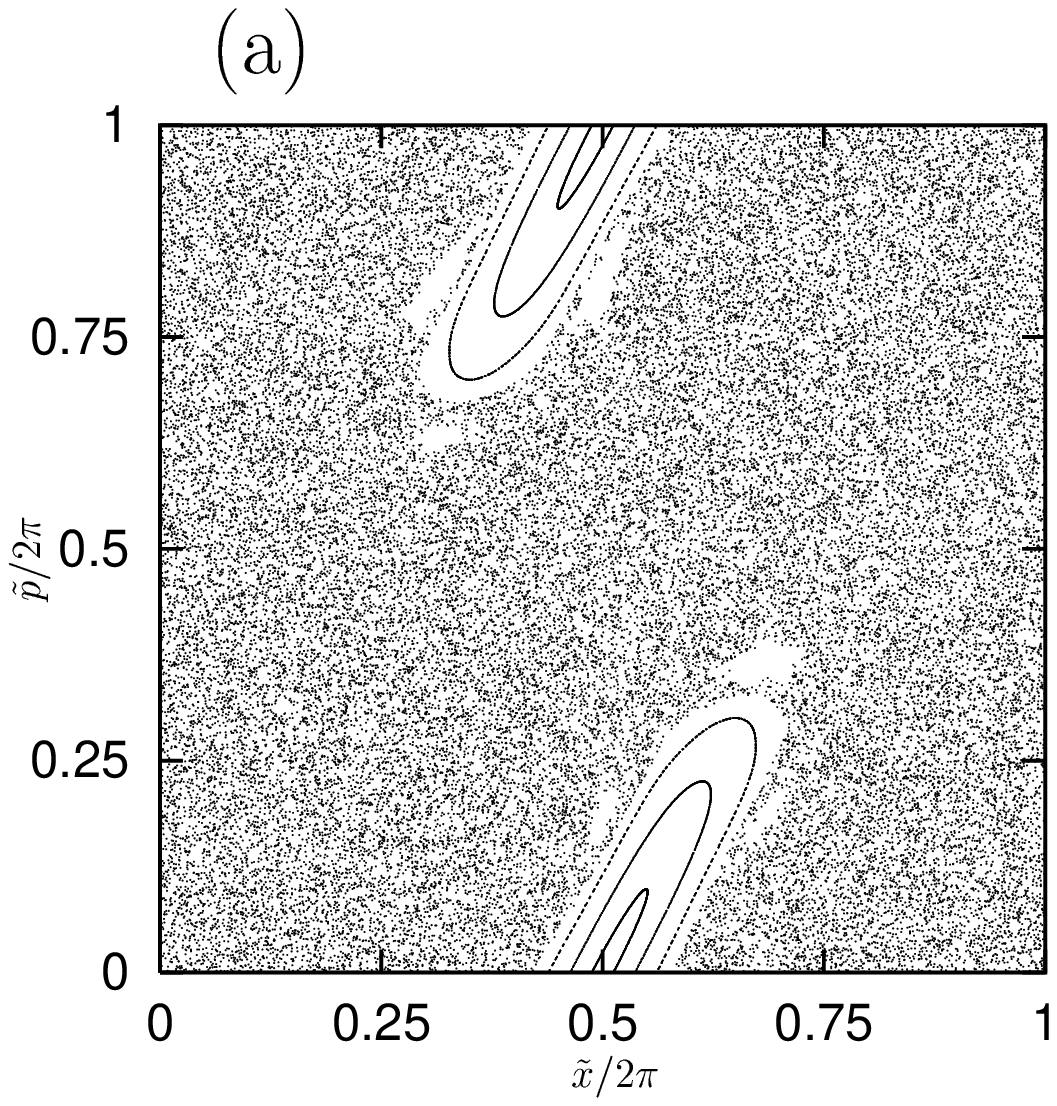,width=8cm}

\vspace{-6.3cm}
\epsfig{file=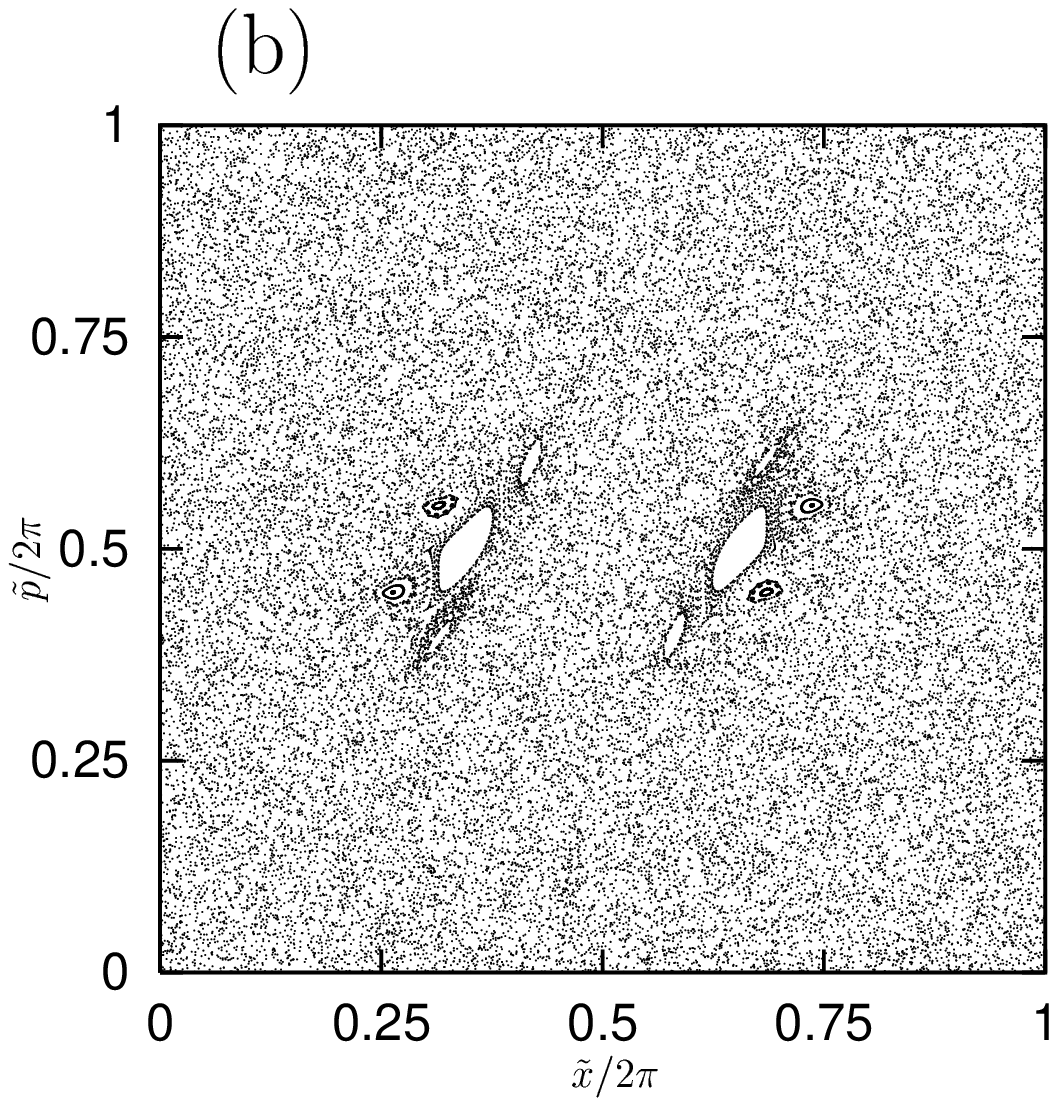,width=8cm}

\vspace{-6.3cm}
\epsfig{file=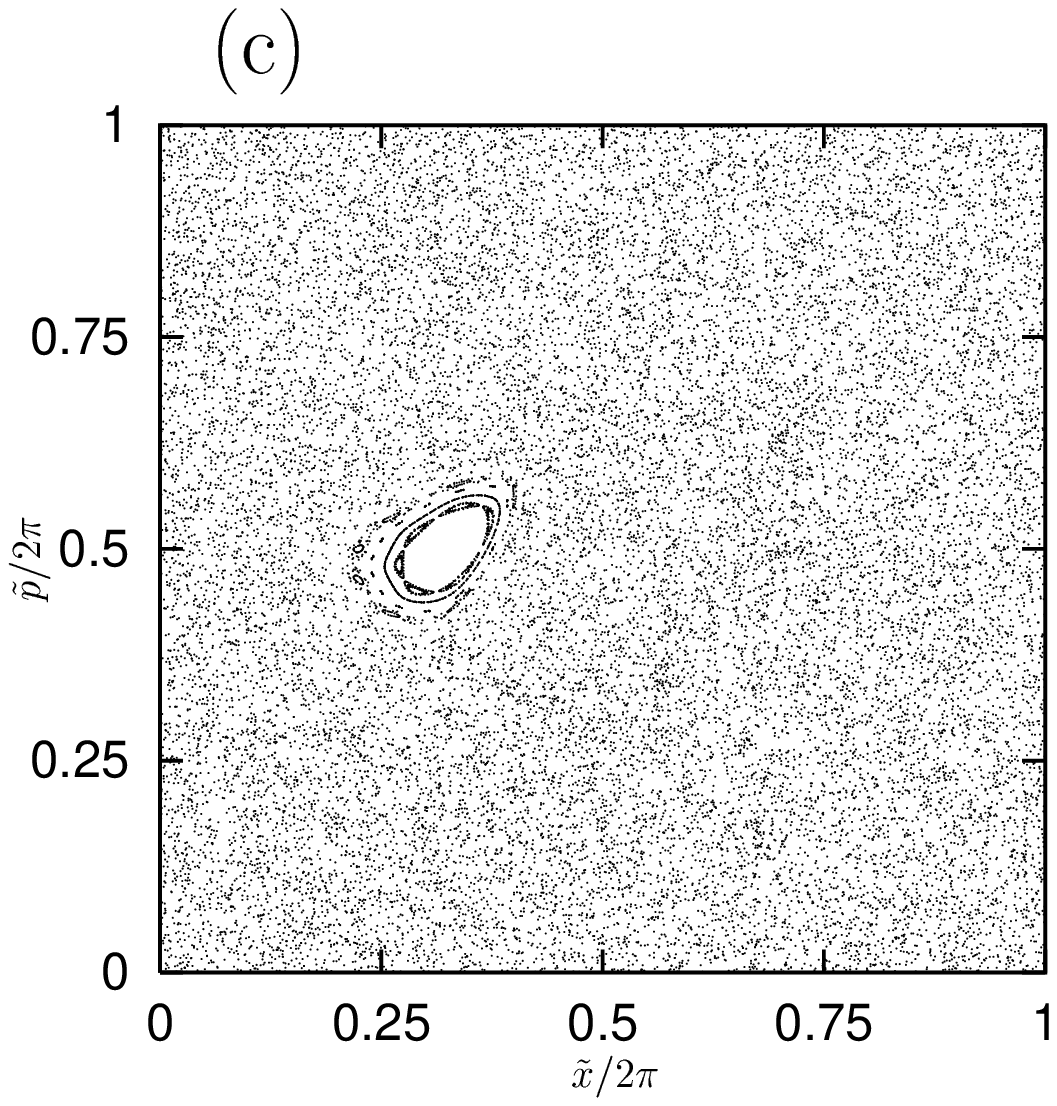,width=8cm}
\end{center}
\vspace{-1.4cm}
\caption{Classical phase space structures of (a) the standard map [$\kappa_{1}=3.8$, $f(n)=1$,
$\kappa_{2}=0$], (b) a modified kicked rotor model [$\kappa_{1}=3.8$, $f(n)=g(n)$, $\kappa_{2}=0$], and (c)
a ratchet accelerator [$\kappa_{1}=3.8$, $f(n)=g(n)$, $\kappa_{2}=0.2$].
All variables are
in dimensionless units.
Note that the regular islands seen in (b) and (c) are
transporting while those in (a) are not.}
\label{clamap}
\end{figure}

To demonstrate the feasibility of our basic idea, 
in this section we present our results using 
some specific computational examples.
Figure \ref{clamap}a displays the classical phase space structure of the standard map with $\kappa_{1}=3.8$, $\kappa_{2}=0$, and $f(n)=1$.
The regular islands in Fig.  \ref{clamap}a  are not transporting since the momentum of the trajectories launched from these regular islands
is bounded and oscillates periodically. Figure \ref{clamap}b shows that if the first kicking field reverses its
potential after every two kicks, then the regular non-transporting islands in the standard map are destroyed and
new regular islands emerge.  A simple computation reveals that
those islands  seen in  Fig. \ref{clamap}b
are transporting. In particular, classical trajectories launched from regular
islands on the left side ($\tilde{x}<\pi$) of the phase space cell
will accelerate, with their $\tilde{p}$ increased  by  $\approx \pi$ after each kick. Due to the spatial symmetry, these islands
always have partners with the opposite transporting property:
trajectories launched from those
islands on the right side ($\tilde{x}>\pi$) of the same phase space cell will accelerate in the opposite direction, with their
$\tilde{p}$ decreased by about $\pi$ after each kick. 
These properties of the transporting regular islands seen in Fig. \ref{clamap}b
indicate that they are associated with the marginal 
stable points $\tilde{p}=\pi$, $\tilde{x}=\pm
\pi/2$ in the case of $\kappa_{1}=\pi$.
Dramatically,
upon
introducing the second kicking field in Eq. (\ref{clamapp}), e.g.,
$\lambda_2/\lambda_1=\kappa_2/\kappa_1=0.2/3.8$,
the transporting islands no longer appear in pairs as in
Fig. \ref{clamap}b. Rather, as seen in Fig. \ref{clamap}c, the spatial symmetry of the phase space structures is clearly broken, but the size of the main
transporting islands that persist remains significant.
Specifically, all the trajectories launched
from the regular structures clearly
seen in  Fig. \ref{clamap}c will still accelerate linearly with time;
the remaining part of the phase space is exclusively occupied by the chaotic sea. That is,
there does not exist similar transporting islands that decreases $\tilde{p}$ in a linear fashion. 

For a classical ensemble that
uniformly covers the entire phase space cell shown in Fig. \ref{clamap}c,  the spatial and temporal periodicity of the system requires that
the average acceleration rate
is zero \cite{schanz00,schanz01}, i.e.,
\begin{equation}
a_{c} A_{c}+ a_{r} A_{r}=0,
\label{vec}
\end{equation}
where $A_{c}$ and $A_{r}$ denote the areas of the chaotic sea and the regular
transporting region, respectively, with their associated average acceleration
rate denoted by
$a_{c}$ and $a_{r}$.
Consider now an initial classical ensemble with  $\tilde{p}=0$ and $\tilde{x}$ uniformly
distributed between $-\infty$ and $+\infty $. Such an ensemble is easy to
simulate both classically and quantum mechanically.
Since this initial ensemble is entirely located in the chaotic sea (see
Fig. \ref{clamap}c),
Eq. (\ref{vec}) predicts that the average acceleration rate of this classical ensemble
is given by $-a_{r}A_{r}/A_{c}$. This also indicates that the larger the transporting island is,
the faster the directed acceleration would be.  In the case of Fig. \ref{clamap}c,
$a_{r}=\pi$, so more trajectories travel to the regime of negative $\tilde{x}$.  
That is, 
even though the average force of the kicking fields is zero and the initial classical ensemble
is both spatially symmetric and time-reversal invariant, 
all the trajectories tend to be accelerated in one direction.
Note that, by changing the strength of the second kicking field
or varying the relative phase between the two kicking fields (e.g., the second kicking field is replaced by
$\lambda_{2} \sin(2x/x_{0}+\beta)\sum_{n}\delta(t/T-n)$, where $\beta$ is a relative phase factor),
one can also
obtain $a_{r}=-\pi$. Hence, the acceleration can be also obtained in the
reverse direction.

\begin{figure}[ht]
\ \vspace{-2cm}
\begin{center}
\epsfig{file=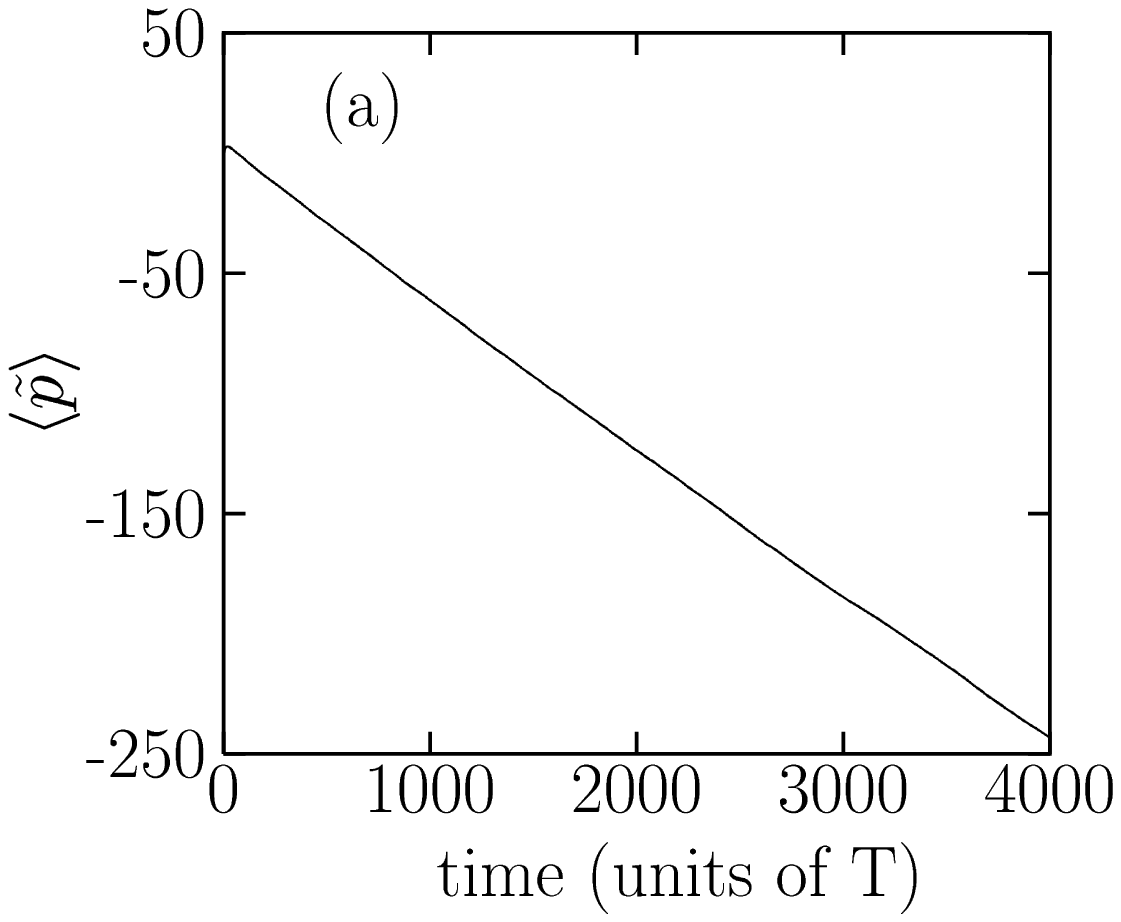,width=11cm}

\vspace{-9cm}
\epsfig{file=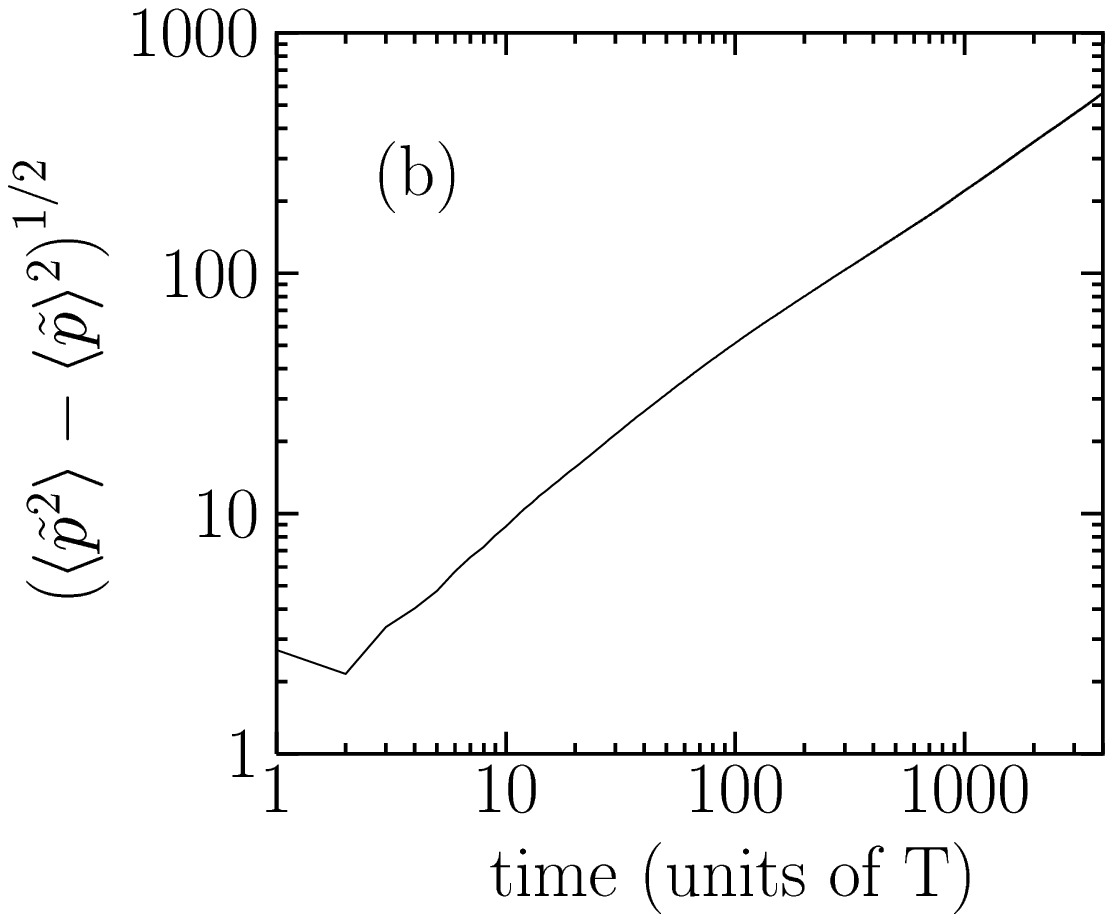,width=11cm}
\end{center}
\vspace{-7cm}
\caption{(a) The linear time dependence of the classical directed current defined
as the ensemble averaged value of the momentum,
with the initial condition $\tilde{p}=0$ and $\tilde{x}$ uniformly
distributed between $-\infty$ and $ +\infty$.
(b) The associated time dependence of the momentum variance in a $\log , \log$ plot.
The system parameters are the same as those in the case of Fig. \ref{clamap}c.
All variables are
in dimensionless units.}
\label{cla-p}
\end{figure}

For an initial classical ensemble with
$\tilde{p}=0$ and $\tilde{x}$ uniformly distributed from 
$-\infty$ to $+\infty$,
Fig. \ref{cla-p}a displays the time dependence of the classical net current $\langle\tilde{p}\rangle$, where $\langle \cdot \rangle$
denotes the ensemble average.
It is indeed seen that the classical net current is in the negative direction, and  accelerates linearly with time. The acceleration rate
is also consistent with Eq. (\ref{vec}).  Figure \ref{cla-p}b displays a $\log-\log$ plot of
the time dependence of the associated momentum variance 
$\Delta_{\tilde{p}}\equiv \left(\langle\tilde{p}^{2}\rangle-\langle\tilde{p}\rangle^{2}\right)^{1/2} $.
A linear fit of the result in  Fig. \ref{cla-p}b gives $\Delta_{\tilde{p}} \sim N^{0.65}$, where $N$ is the number of kicks.
This confirms that the diffusion dynamics is anomalous and
immediately gives $ \Delta_{\tilde{p}}/\langle\tilde{p}\rangle \sim 1/N^{0.35}$. Hence, as far as the first and second
order statistical moments are concerned,  
the momentum fluctuations of the classical ensemble become less and less
important as time increases. In this sense, for sufficiently large times,
almost all trajectories in the classical
ensemble are accelerated in the negative direction. 

\begin{figure}[ht]
\ \vspace{-2cm}
\begin{center}
\vspace{2cm}
\epsfig{file=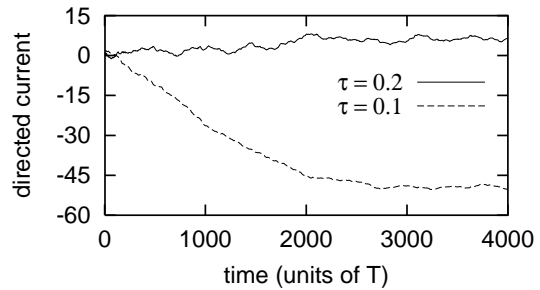,width=4cm,angle=270}
\end{center}
\caption{The time dependence of the quantum dimensionless directed current defined as  the expectation value of $\hat{\tilde{p}}$
for two values of the effective Planck constant $\tau$,
with the initial condition given by $|0\rangle$ and other parameters are
the same as in Fig. \ref{clamap}c.
Results here should be compared with those in  Fig. \ref{clamap}c.
}
\label{q-current}
\end{figure}

\section{Quantum Results}
How then does this ratchet accelerator manifests itself in the corresponding quantum system and
how small the effective Planck constant should be in order to observe the analogous ratchet effects in the quantum dynamics?  To carry
out
the quantum calculations we impose a periodic boundary condition on the system, i.e.,
$|\psi(\tilde{x})\rangle=|\psi(\tilde{x}+2\pi)\rangle$, where $|\psi\rangle$ is the wavefunction of the system. This
periodic boundary condition suggests that
$\tilde{x}$ is now to be understood as an angle variable, and  
is consistent with the fact that the classical dynamics is invariant if $\tilde{x}\rightarrow \tilde{x}+2\pi$.
Further,
our previous work \cite{gongpre1} suggests that
this choice of boundary conditions should not affect the essence of the quantum dynamics.

Figure \ref{q-current} shows the net quantum current, 
i.e., the expectation value
of $\hat{\tilde{p}}$ for two values of the effective Planck constant $\tau$, with the initial state given by $|\psi(t=0)\rangle=|0\rangle$.
Comparing the results here with that in Fig. \ref{cla-p}a, one
sees that the magnitude of the quantum directed current can be much smaller than that of the
classical directed current.  Interestingly,
the directed current in the quantum case
displays much more complicated behavior
than the linear acceleration seen classically (some slow  oscillatory behavior of
the quantum current over much longer
time scales is also observed, but should be
of much less
experimental interest).
This can be qualitatively explained in terms 
of a combination of a dynamical localization effect \cite{casatibook} and
the quantum tunneling between the chaotic sea and the transporting islands
\cite{gongpre1,schanz02}. 
In particular, since the dynamical 
localization in quantum delta-kicked systems
always suppresses the diffusion, 
the quantum directed current is not expected to grow all the time.
More importantly,
since the chaotic sea and the transporting islands have opposite acceleration signs [see Eq. (\ref{vec})], the quantum
tunneling between them \cite{gongpre1} necessarily decreases the directed current.
Indeed, 
in the case of $\tau=0.2$, one sees that for most of the time the quantum 
current is
in the opposite direction of the classical current.
Note that,
unlike a previous observation of tunneling-induced current reversal in quantum Brownian motion \cite{reimann},
here the tunneling is between  a  chaotic sea and a regular region and the
system is totally isolated from the environment. 

The results in Fig. \ref{q-current} also suggest that $\tau=0.1$ is
small enough to observe the signature of
a ratchet accelerator in a quantum delta-kicked  system. 
For example, the quantum current  for $\tau=0.1$ does increase almost
linearly for as many as 2000 kicks, with the average increase rate
about three times smaller than in the classical case.
This means that for experiments using cold sodium atoms \cite{raizenetc} the directed and relatively large
quantum current obtained with 2000 kicks would
correspond to an average drift velocity on the order of 1 m/s.
Note however, this value of $\tau$ is about one order
of magnitude smaller than what is being
examined in atom optics experiments \cite{raizenetc},
but may be achievable in the future.  

\begin{figure}[ht]
\vspace{-3cm}
\begin{center}
\vspace{2cm}
\epsfig{file=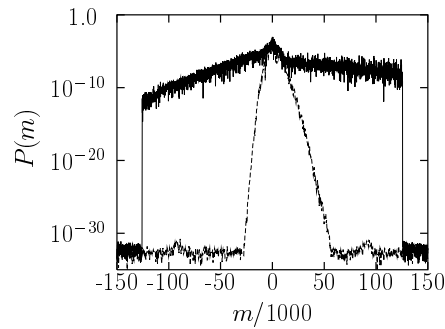,width=10cm}
\end{center}
\vspace{-6.5cm}
\caption{The probability $P(m)$
of finding the quantum system in state $|m\rangle$ after 4000 kicks, with the initial state given by $|0\rangle$.
Solid line is for $\tau=0.1$ and
dashed line is for  $\tau=0.2$. Other parameters are the same as those in the case of Fig. \ref{clamap}c.
}
\label{pdis}
\end{figure}

It is also interesting to examine the lineshape of dynamical localization with the spatial symmetry of the transporting regular
islands broken by the second kicking field.
Figure \ref{pdis} displays the associated distribution function $P(m)$ after 4000 kicks
for the same two values of $\tau$.
It is clearly seen that for both cases $P(m)$ is  asymmetric. 
Due to the stickiness of the boundary between the transporting islands and the chaotic sea and the associated quantum tunneling, 
the probability of finding the system in state $|m\rangle$ with a large and positive $m$ can be much higher than
that of finding the system in state $|-m\rangle$.  This does not contradict the previous observation
that after an initial stage the quantum current for $\tau=0.1$ is negative, because the tail of the statistical distribution function
$P(m)$ has little effect on the net current, the 
statistical moment of the lowest order.  The huge difference in $P(m)$ between the two cases shown in Fig. \ref{pdis}
with slightly different values of $\tau$ is also
noteworthy.  This extreme sensitivity of $P(m)$ to the value of $\tau$
is more or less related to the fact that the values of $\tau$ we
consider are quite comparable to the area of the main
transporting island seen in Fig. \ref{clamap}c.

\section{Concluding Remarks}

Using a simple model Hamiltonian system we have demonstrated a new
ratchet effect by considering directed classical anomalous
diffusion. We have shown that it is possible to achieve directed acceleration with a fixed acceleration rate
with two kicking fields that have zero average gradient.
This is the case even though the initial classical ensemble is
both spatially symmetric and time-reversal invariant.  We have also presented some
computational examples of the quantum dynamics with different values
of the effective Planck constant. It is found that quantum coherence effects may induce a reversal of 
directed current. It is also found that directed  quantum current that accelerates linearly with time
without using a biased external field should also be
achievable in a quantum system
that is still far from the classical limit.

We have previously shown that quantum anomalous diffusion in a simple delta-kicked system
can be faster than the underlying classical anomalous diffusion
\cite{gongpre1}. However, the spatial symmetry therein always gives a zero momentum expectation value and therefore
one necessarily considers the momentum  variance in order to make a meaningful quantum-classical comparison. By contrast, our
ratchet accelerator model here allows for a study of the quantum-classical differences in the  momentum expectation value itself.
As such, the role of the quantum tunneling between a chaotic sea and transporting regular islands becomes even
more clear: while it can enhance the increase rate of the momentum variance
\cite{gongpre1}, it suppresses the net acceleration rate
of the average momentum in our ratchet accelerator model.

It would be of great interest to consider a molecular realization of the ratchet accelerator, where the variable $x$ is understood
as an angle variable and $p$ is understood as an angular momentum variable.
Since a microwave field \cite{fishman,gongjcp} will create a kicking potential of the form $\cos(x)$
and an off-resonance laser field \cite{averbukh,averbukh2} can create a kicking potential of the form  $\cos^{2}(x)$,
a diatomic subject to both microwave and off-resonance laser fields, with their polarization direction perpendicular to one another,
is a possible candidate for realizing 
the ratchet accelerator we propose.  However, since in this case the diatomic should be at least described by a three-dimensional rigid rotor,
which differs from
the planetary kicked rotor in a number of aspects \cite{gongjcp},  detailed calculations should be carried out to guide the experimental
studies.
Nevertheless, we believe that this ratchet accelerator model
may be of importance for manipulating the rotational motion of diatomics and may provide a means of orienting molecules
while accelerating rotation without using
a biased external field.

This work was supported by the U.S. Office of Naval Research, Photonics Research
Ontario, and the
Natural Sciences and Engineering Research Council of Canada.


\begin{thebibliography}{100}
\bibitem[*]{bylineone} Current Address: Department of Chemistry and The James
Franck Institute,
University of Chicago, Chicago,
IL 60637.
\bibitem{reimann02} P. Reimann, Phys. Rep. {\bf 361}, 57 (2002).
\bibitem{astumian} R.D. Astumian and P. H\"{a}nggi, Phys. Today {\bf 55}, 33 (2002).
\bibitem{jung96} P. Jung, J.G. Kissner, and P. H\"{a}nggi, \prl{\bf 76}, 3436 (1996).
\bibitem{mateos} J.L. Mateos, \prl{\bf 84}, 258 (2000).
\bibitem{porto} M. Porto, M. Urbakh, and J. Klafter, \prl{\bf 85}, 491 (2000).
\bibitem{flach} S. Flach, O. Yevtushenko, and Y. Zolotaryuk, \prl{\bf 84}, 2358 (2000).
\bibitem{schanz00} T. Dittrich, R. Ketzmerick, M.F. Otto, and H. Schanz, Ann. Phys. (Berlin), {\bf 9},  755 (2000).
\bibitem{schanz01} H. Schanz, M.F. Otto, R. Ketzmerick, and T. Dittrich, \prl{\bf 87}, 070601 (2001).  Transporting trajectories can be transporting in either
momentum or position, whereas accelerating trajectories are transporting in
momentum and hence accelerate.
\bibitem{cheon02} T. Cheon, P. Exner, and P. Seba,
 J. Phys. Soc. Japan {\bf 72},  1087 (2003).
\bibitem{denisov02} S. Denisov, J. Klafter, and M. Urbakh, \pre{\bf 66}, 046203 (2002).
\bibitem{monteriro02} T.S. Monteiro, P.A. Dando, N.A.C.  Hutchings, and M.R. Isherwood, \prl{\bf 89}, 194102 (2002).
\bibitem{jonckheere03} T. Jonckheere, M.R. Isherwood, and T.S. Monteiro, Phys. Rev. Lett. {\bf
91}, 253003 (2003).
\bibitem{jones03} P.H. Jones, M. Goonasekera, H.E. Saunders-Singer, and D.R. Meacher, preprint quant-ph 0309149 (2003).
\bibitem{casatibook} G. Casati and B. Chirikov, {\it Quantum chaos:
 between order
  and disorder} (Cambridge University Press, New York, 1995).
  \bibitem{raizenetc}
  F.L. Moore, J.C. Robinson, C.F. Bharucha, B. Sundaram, and M.G. Raizen,
   \prl{\bf 75}, 4598 (1995);
   H. Ammann, R. Gray, I. Shvarchuck, and N. Christensen,
    \prl{\bf 80}, 4111 (1998);
    J. Ringot, P. Szriftgiser, and J.C. Garreau, D. Delande,  \prl{\bf 85}, 2741 (2000);
     M.B. d'Arcy, R.M. Godun, M.K. Oberthaler, D. Cassettari, and G. S. Summy,
      \prl{\bf 87},
	074102 (2001).
\bibitem{fishmanprl} S. Fishman, D.R. Grempel, and R.E.  Prange,
\prl{\bf 49}, 509 (1982).
\bibitem{casatiprl}G. Benenti, G. Casati, I. Guarneri, and M.
      Terraneo, \prl{\bf 87}, 014101 (2001).
   \bibitem{fishman} R. Bl\"{u}mel, S. Fishman, and U. Smilansky,
 J. Chem. Phys. {\bf 84},
     2604 (1986).
   \bibitem{averbukh}
   I. Sh. Averbukh and R. Arvieu, \prl{\bf 87}, 163601 (2001).
   \bibitem{machholm} M. Machholm and N.E. Henriksen, \prl{\bf 87},
   193001 (2001).
   \bibitem{averbukh2} M. Leibscher, I.Sh. Averbukh, and H. Rabitz,
   \prl{\bf 90}, 213001 (2003).
 \bibitem{facchi}P. Facchi, S. Pascazio, and A. Scardicchio, \prl{\bf 83}, 61 (1999).
 \bibitem{georgeot}B. Georgeot and D.L. Shepelyansky, \prl{\bf 86}, 2890
 (2001).
\bibitem{kurizki} G. Kurizki, M. Shapiro, and P. Brumer, \prb{\bf 39}, 3435 (1989);
M. Shapiro and P. Brumer, {\it Principles of
the Quantum Control of Molecular Processes} (John Wiley, New York, 2003).
 \bibitem{hache}
R.  Atanasov, A. Hache, J.L.P. Hughes, H.M. van Driel, and J.E. Sipe,  \prl{\bf 76}, 1703 (1996).
\bibitem{bhat} R.D.R. Bhat and J.E. Sipe, \prl{\bf 85}, 5432 (2000).
\bibitem{dupont}E. Dupont, P.B. Corkum, H.C. Liu, M. Buchanan, and Z.R. Wasilewski, \prl{\bf 74}, 3596 (1995).
\bibitem{hache2}
A. Hache, Y. Kostoulas, R. Atanasov, J.L.P. Hughes, J.E. Sipe, and H.M.  van Driel, \prl{\bf 78}, 306 (1997).
\bibitem{stevens}
 M.J. Stevens, A.L. Smirl, R.D.R. Bhat, A. Najmaie, J.E. Sipe, and H.M. van Driel,
 \prl{\bf 90}, 136603 (2003).
\bibitem{gongpre1}J. Gong, H.J. W\"{o}rner, and P. Brumer, \pre{\bf 68}, 026209 (2003).
\bibitem{gongpre2}J. Gong, H.J. W\"{o}rner, and P. Brumer, \pre{\bf 68}, 056202 (2003).
\bibitem{zaslavsky02}A. Iomin, S. Fishman, and G.M. Zaslavsky, \pre{\bf 65}, 036215 
(2002).
\bibitem{grigolini} R. Roncaglia, L. Bonci, B.J. West, and P. Grigolini, \pre{\bf 51}, 5524
(1995).
\bibitem{denisov} S. Denisov and S. Flach, \pre{\bf 64}, 056236 (2001).
\bibitem{denisov2}
S. Denisov, J. Klafter, M. Urbakh, and S. Flach,  Physica D {\bf 170}, 131
(2002).  
\bibitem{denisov3}
S. Denisov, S. Flach, A.A. Ovchinnikov, O. Yevtushenko, and Y.
Zolotaryuk  \pre{\bf  66}, 041104 (2002).
\bibitem{schanz02} The quantum tunneling between the chaotic sea and the transporting island implies that
the Floquet state of the kicked quantum system can simultaneously occupy these classical
phase space structures, a counter-intuitive result first demonstrated in L. Hufnagel, R. Ketzmerick, M.F. Otto, and H. Schanz,
\prl{\bf 89}, 154101 (2002). 
\bibitem{reimann} P. Reimann, M. Grifoni, and P. H\"{a}nggi, \prl{\bf 79}, 10 (1997).
\bibitem{gongjcp} J. Gong and P. Brumer, \jcp{\bf 115}, 3590 (2001).
\end{thebibliography}
\end{document}